\documentclass[conference]{IEEEtran}
\usepackage{enumitem}

\usepackage[T1]{fontenc}
\usepackage{color}

\usepackage{cite}

\ifCLASSINFOpdf
  \usepackage[pdftex]{graphicx}
\else
\fi

\usepackage{amsmath}
\usepackage[cmintegrals]{newtxmath}
\def\ket#1{\mathinner{|{#1}\rangle}}
\def\braket#1{\mathinner{\langle{#1}\rangle}}

\def\Ket#1{\left|#1\right\rangle}
\def\Braket#1{\left\langle#1\right\rangle}

\usepackage{array}

\usepackage{url}

\makeatletter
\newcommand*{\Rom}[1]{\expandafter\@slowromancap\romannumeral #1@}
\newcommand*{\rom}[1]{\expandafter\romannumeral #1}
\makeatother

\usepackage[normalem]{ulem}

\begin{document}

%
\title{Multiantenna Quantum Backscatter Communications}
%
%
%
\author{\IEEEauthorblockN{Riku J\"{a}ntti\IEEEauthorrefmark{1},
Roberto Di Candia\IEEEauthorrefmark{2},
Ruifeng Duan\IEEEauthorrefmark{1}, and  
Kalle Ruttik\IEEEauthorrefmark{1}
}
\IEEEauthorblockA{\IEEEauthorrefmark{1}Department of Communications and Networking, Aalto University, Espoo, 02150 Finland.\\ Email: \{riku.jantti; ruifeng.duan; kalle.ruttik\}@aalto.fi}
\IEEEauthorblockA{\IEEEauthorrefmark{2}Dahlem Center for Complex Quantum Systems, Freie Universit\"{a}t Berlin, 14195 Berlin, Germany.\\
Email: rob.dicandia@gmail.com}
}


%



\maketitle

\begin{abstract}
Quantum  illumination  (QI)  is  a  revolutionary  photonic  quantum  sensing  paradigm  that  enhances  the  sensitivity of photodetection in noisy and lossy environments. The QI concept has been recently used to propose a quantum backscatter communication (QBC), with the aim of increasing the receiver sensitivity beyond the limits of its classical counterpart. One of the practical challenges in microwave QI is the slow rate at which the entangled microwave modes can be generated. Here, we propose to mitigate this problem by using a multiple-input multiple-output antenna system to synthetically increase the number of efficiently-distinguishable modes in the QBC context.


\end{abstract}


%
\IEEEpeerreviewmaketitle

\section{Introduction}
Backscatter of radio waves from an object is the subject of active studies since the development of radars in the 1930s, and the use of backscattered radio for communications since H. Stockman's work in 1948~\cite{Stockman1948}. Backscatter communication (BC) is widely used in radio frequency (RF) tags [e.g., for RF identification]~\cite{Chawla2007overviewRFID}, and it bears close resemblance with the radar. In fact, both systems can be described with the so-called radar equations, which state that the propagation is inversely proportional to the fourth power of the distance, and that the power reflected from the target can be described with the radar cross section (RCS) parameter~\cite{Brandsema2017theoretical}. In the case of BC, the RCS needs to be determined for the RF tag antenna~\cite{Yen2007radar}. 

The concept of quantum illumination (QI) was introduced by S. Lloyd in 2008~\cite{Lloyd2008, Tan08}. It consists in using entangled photons to increase the success probability of detecting a low-reflectivity object in a noisy and lossy environment. The application in the microwave regime~\cite{Barzanjeh2015} was proposed afterwards, and it paved the way to a prototype of quantum radar. While the quantum radar model is based on detecting the reflected signal of a object possibly present in a target region, which can be thought as an on-off keying (OOK), a typical BC protocol involves different phase and amplitude modulations.
The quantum backscatter communication (QBC) systems bear close resemblance to the the quantum radar, and are characterized by the quantum RCS (QRCS), which relates the intensity of photons scattered from the target to the intensity of the incident photons~\cite{Liu2014}. 

Practical applications of QI are limited by the rate $N_s$ at which the microwave illumination device can generate $M$ independent entangled mode pairs~\cite{Lanzagorta2016}. The parameter $M=WT$ depends on the bandwidth $W$ and pulse duration $T$, which in QBCs is related to the symbol duration of the backscatter device. In order for a quantum protocol to achieve small error probabilities, the number of modes must be of the order $M\sim 10^9$, while in the radar band we can achieve at most $M\sim 10^5$~\cite{Lanzagorta2016}. While in a classical setup this problem can be overcome by increasing the input signal power, in QI this would lead to the lost of the quantum advantage for several reasons. First, the receiver would tend to perform as a classical receiver. Second, a quantum signal with several average numbers of photons is challenging to create, due to the increasing thermal effect arising in this regime. In~\cite{Lanzagorta2016}, so-called "virtual" modes have been introduced in order to improve the performance without increasing $W$ and $T$. The proposed system consists of several parallel microwave illumination devices with the aim of detecting the same target. The limit of this approach consists in the presence of a mutual interference between the simultaneous transmissions, resulting in an increasing of the background noise and a higher detection error probability. 

In this paper, we propose a beam-splitter based precoding and receiver beamforming, able to perform parallel QI of a target without interference. The target application is a QBC where the properties of the target, being the communicating device, are known in advance. It is known that in classical wireless backscatter channel, the cluttering objects around the target antenna give rise to fast fading which can be mitigated by using multiple antennas~\cite{griffin2008gains}. We show how multiple antennas can be utilized to obtain spatial diversity and mitigate the impact of fading also in the quantum setting. 

This paper is organized as follows. In Section~\ref{sec:chann_model}, we describe how a multiple-input multiple-output (MIMO) channel can be modeled using beam-splitters. Section~\ref{sec:Qreceivers} introduces the Quantum receivers and discusses quantum protocols for demodulating binary phase shift keying (BPSK) symbols. Finally, section~IV concludes the paper.

\section{Channel model}\label{sec:chann_model}

In radio communication, the transmit antenna oscillating with angular frequency $\omega$ generates an electromagnetic field, which in the low photon-number regime is quantized. An introduction of quantum electrodynamics theory is out of the scope of this paper. We simply note that the free Hamiltonian of quantized electromagnetic field has the same form as the Hamiltonian of quantum harmonic oscillator  $H=\hbar\omega\left(\hat{a}^\dagger\hat{a}+1/2\right)$, where $\hbar$ denotes the reduced Planck constant, and the operators $\hat {a}$ and $\hat {a}^\dag$ are called annihilation and creation operators of the harmonic oscillator, respectively~\cite{Helstrom1976}. Their action on the eigenvectors of the Hamiltonian $H$ is given by $\hat {a}|n\rangle=\sqrt{n}|n-1\rangle$ and $\hat {a}^\dag |n\rangle=\sqrt{n+1}|n+1\rangle$~\footnote{The basis $\{|n\rangle\}_{n=0}^\infty$ is usually denoted as Fock basis.}. Coherent states are the eigenstates of the operator $\hat a$. In this context, they are referred to as classical states, as the statistics of their measurements resembles the one of the classical signals.  The canonical position and momentum-like operators define the in-phase and quadrature components given by $\hat x= (\hat {a}^\dag+\hat {a})/\sqrt{2}$ and $\hat p=i(\hat {a}^\dag-\hat {a})/\sqrt{2}$, respectively, where $i=\sqrt{-1}$ denotes the imaginary number.


\subsection{Single-input signal-output case}
A beam splitter can be modeled in the Heisenberg picture by a 2x2 unitary matrix~\cite{leonhardt2003quantum}. This operation describes the effect of the channel on the incident modes which results in the outgoing modes. Consider a single-input signal-output (SISO) BC system having a single transmit antenna, a single BC antenna, and a single receive antenna. The channel can be modeled as a simple beam splitter~\cite{leonhardt2003quantum,Heras2016}:

\begin{equation}\label{eq1}
\begin{pmatrix} \hat{a}_R\\ \hat{a}_Z' \end{pmatrix}=
\underbrace{\begin{bmatrix} 
\sqrt{\eta}e^{-i\phi} & \sqrt{1-\eta}\\
-\sqrt{1-\eta} & \sqrt{\eta}e^{i\phi}
\end{bmatrix}}_{\boldsymbol{B}_{\eta,\phi}}
\begin{pmatrix}\hat{a}_S\\ \hat{a}_Z \end{pmatrix},
\end{equation}
where the parameter $\eta$ denotes the \emph{round trip transmissivity} (RTT), $\hat{a}_S$ is the transmitted signal mode, $\hat{a}_Z$ is the thermal mode modeling the bright thermal environment, $\hat{a}_R$ is the the received mode, and  $\hat{a}_Z'$ is the mode of the lost photons.  
In Eq.~\eqref{eq1}, the phase shift $\phi$ of the channel depends on the communication distance $R$, and the phase shift ${\phi=2\pi R /\mathsf{c} + \varphi}$ caused by the backscatter antenna depends on a phase $\varphi$ and on the speed of light $\mathsf{c}$. In particular, the backscatter antenna can control the phase $\varphi$, where the communication is encoded. The RTT $\eta$ can be represented as:
\begin{equation}\label{eq:eta_sigma}
    \eta=\frac{G^2\mathsf{c}^2\sigma_Q}{16\pi \omega^2 R_t^2R_r^2}, 
\end{equation}
which depends on the antenna gain $G$ of the reader, on the distance from transmitter to the tag $R_t$, on the distance from tag to receiver $R_r$, and on the operating frequency $\omega$.
The quantity $\sigma_Q={\Braket{\hat{I}_s}}/{\Braket{\hat{I}_i}}$ defines the QRCS~\cite{Liu2014}. Here, $\Braket{\hat{I}_s}$ denotes the intensity of the reflected signal, and $\Braket{\hat{I}_i}$ is the intensity of the incident signal. In the case of QBCs, the QRCS depends on the properties of the BC antenna, and $\sigma_Q<1$ due to the modulation losses at the antenna. We assume that $\sigma_Q$ remains constant in the time window $T$, and that the communication takes place by only controlling the phase $\varphi$. Moreover, in practice we have $\eta\ll1$ due to the large communication distance $R_t+R_r\gg1$.

\subsection{Two-by-two MIMO case}
\begin{figure}[t]
	\centering
	\includegraphics[width=\columnwidth]{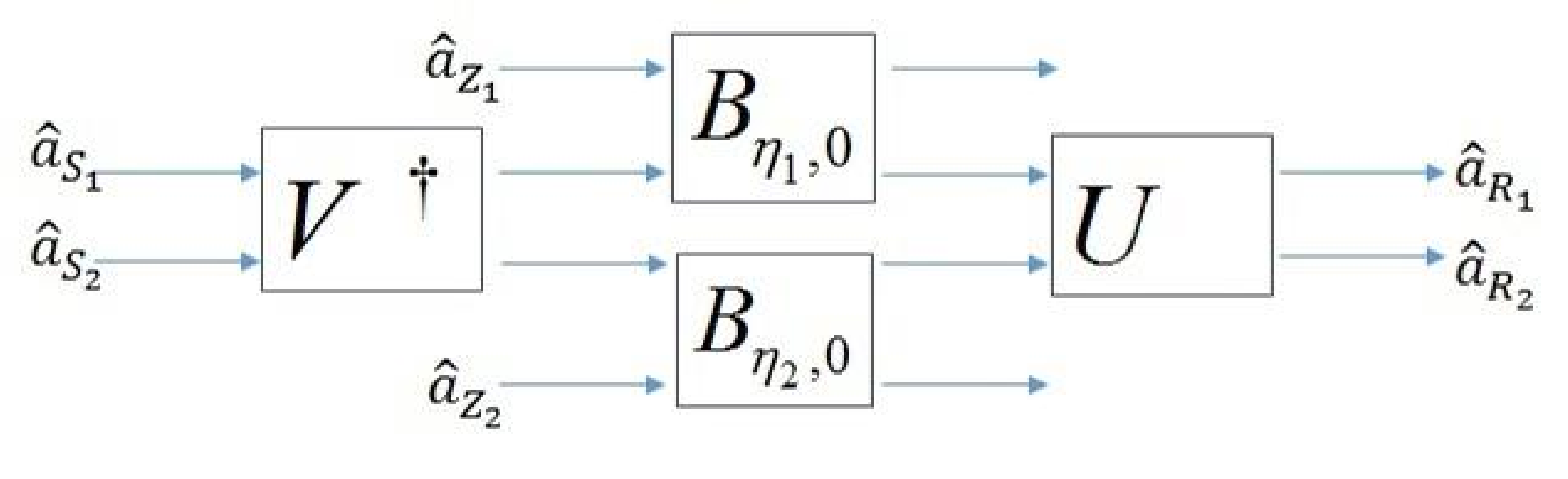}
	\caption{Beam-splitter model for a two-by-two MIMO channel.} 
	\label{fig:channel}
\end{figure}

Let us consider a system with two transmitting and two receiving antennas. We assume that all the antennas perform BPSK modulation using the same symbol $x=e^{-i\varphi}$. Without loss of generality, we set $x=1$ in this section.
Let $\hat{\boldsymbol{a}}_R=(\hat{a}_{R,1}^\dag,\hat{a}_{R,2}^\dag)^\dagger$, $\hat{\boldsymbol{a}}_S=(\hat{a}_{S,1}^\dag,\hat{a}_{S,2}^\dag)^\dagger$, and $\hat{\boldsymbol{a}}_Z=(\hat{a}_{Z,1}^\dag,\hat{a}_{Z,2}^\dag)^\dag$ denote the vectors of received, transmitted and thermal modes, respectively.~\footnote{The superscript ${}^\dag$ denotes the transpose operation.} The input-output relations of the channel are described by a $2\times 2$ matrix $\boldsymbol{H}=[h_{nm}]$, where the elements of the matrix $\{h_{nm}\}_{n,m=1}^2$ represents the complex probability amplitudes of receiving a photon transmitted from antenna $n$ to the receive antenna $m$. Here, we assume that the channel act passively on the $\hat{\boldsymbol{a}}_{S,Z}$ modes, so that the outputs $\hat{\boldsymbol{a}}_R$ do not depend on the conjugate fields $\hat{\boldsymbol{a}}_{S,Z}^\dag$. The singular value decomposition (SVD) of the channel matrix ${\boldsymbol{H}}$ can be written  as  ${\boldsymbol{H}}=\boldsymbol{U}{\Sigma}\boldsymbol{V}^\dagger$, where $\boldsymbol{U}$ and $\boldsymbol{V}$ are unitary matrices and $\boldsymbol{\Sigma}=\mathrm{diag}\left\{\sqrt{\eta}_1,\sqrt{\eta}_2 \right\}$ are the singular values of ${\boldsymbol{H}}$. We can model the wireless channel with four beam-splitters: Input beam-splitter $\boldsymbol{V}^\dagger$ mixes the two input modes and yields two outputs as shown in Fig. \ref{fig:channel}. Then, the first output is mixed with a vacuum state in beam-splitter $\boldsymbol{B}_{\eta_1,0}$, and the second output is mixed with a vacuum state in beam-splitter $\boldsymbol{B}_{\eta_2,0}$. One output of each of these two mixers $\boldsymbol{B}_{\eta_k,\varphi}$ ($k=1,2$), represents the absorbed and lost photons. The other outputs are mixed with output beam-splitter $\boldsymbol{U}$. 
The resulting model becomes
\begin{equation} \label{eq:SVD}
    \hat{\boldsymbol{a}}_R=\boldsymbol{U\Sigma V}^\dagger \hat{\boldsymbol{a}}_S + \boldsymbol{US}\hat{\boldsymbol{a}}_Z,
\end{equation}
where  $\boldsymbol{S}=\mathrm{diag}\{\sqrt{1-\eta_1},\sqrt{1-\eta_2}\}$.

\begin{figure*}[t]
	\centering
	\includegraphics[width=0.85\textwidth]{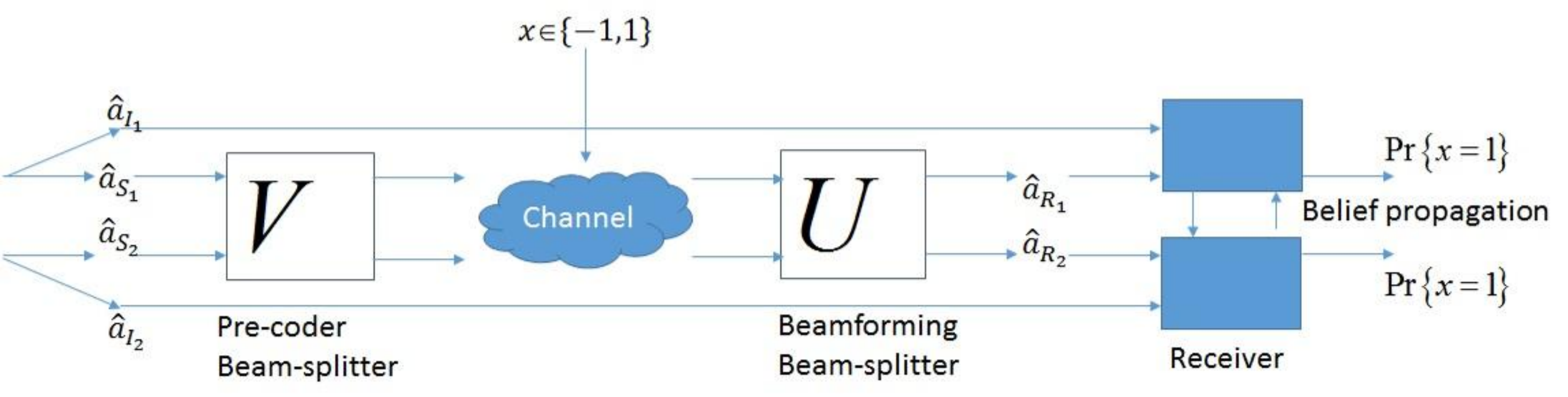}
	\caption{E-MIMO system for a two-by-two channel matrix.} 
	\label{fig:system}
\end{figure*}

In classical wireless communications it has been noted that the MIMO backscatter channel~\cite{griffin2008gains} resembles the MIMO keyhole channel~\cite{chizhik2002_keyholes}. In a keyhole channel all radio paths from the transmitter antennas to the receiver antennas need to pass through a single point, a key hole. In BC channel, the BC tag antennas are kind of keyholes except they modulate the signal that passes through them~\cite{Duan2017ARS}.

Assume that the transmit and receive antennas form an array with antenna spacing $\Delta$. Let $\theta_{t,k}$ denote the angle between the linear transmit array and the backscatter antenna $k$, and  $\theta_{r,k}$ denote the angle between the linear receiver array and the backscatter antenna $k$. We define that  $\Omega_{t,k}=\cos\theta_{t,k}$, $\Omega_{r,k}=\cos\theta_{r,k}$, and $\boldsymbol{e}(\Omega)^\dagger=\begin{pmatrix}1 & e^{i2\pi\Delta\Omega}\end{pmatrix}/{\sqrt{2}}$. The channel coefficients with two BC antennas can be written as
\begin{equation}
    \boldsymbol{H}=\sum_{k=1}^2 \sqrt{\eta_k'}e^{-i\phi_k'}\boldsymbol{e}(\Omega_{r,k})\boldsymbol{e}(\Omega_{t,k})^\dagger,
\end{equation}
where the term $\sqrt{\eta_k'}e^{i\phi_k'}$ is related to the channel transmissivity, the attenuation, and the phase shift. The above matrix is of full rank for $2\times 2$ MIMO scenario, if the two backscatter  antennas are separated in space such that $\Omega_{t,1}\neq \Omega_{t,2} \mod \Delta^{-1}$ and $\Omega_{r,1}\neq \Omega_{r,2} \mod \Delta^{-1}$. If the BC antennas form an array, this condition is likely not to be satisfied and the rank of the channel matrix will be 1. However, in this case we can still benefit from multiple antennas through the beamforming gain.

\subsection{$N_r\times N_t$ MIMO case}
Let us now consider the case in which the transmitter has $N_t$ antennas and the receiver has $N_r$ antennas. We note that (\ref{eq:SVD}) holds also in $N_r\times N_t$ MIMO. Now 
$\boldsymbol{V}^\dagger$ is a $N_t\times N_t$ matrix, $\boldsymbol{U}$ is a $N_r\times N_r$ matrix, and $\Sigma$ is a $N_r\times N_t$ matrix of the form
\begin{equation}
    \boldsymbol{\Sigma}=\begin{bmatrix} \textrm{diag}\left\{\sqrt{\eta}_1,\cdots,\sqrt{\eta}_r\right\} & \boldsymbol{0}_{r\times (N_t-r)}\\ \boldsymbol{0}_{(N_r-r)\times r} &\boldsymbol{0}_{(N_r-r)\times (N_t-r)} \end{bmatrix},
\end{equation}
where $r$ denotes the rank of the matrix $\boldsymbol{H}$ and $\boldsymbol{0}_{m\times n}$ is $m\times n$ matrix full of zeros. The form of the noise matrix does not change and is given by $\boldsymbol{S}=\mathsf{diag}\{\sqrt{1-\eta_1},\cdots,\sqrt{1-\eta_{N_r}}\}$. In \cite{clements2016optimal}, it is shown that any $N\times N$ unitary matrix $\boldsymbol{U}$ can be decomposed to a regular mesh of interconnected two port beam-splitters. Consequently, $\boldsymbol{U}$ and $\boldsymbol{V}^\dagger$ can be decomposed into a grid of beam-splitters.

We note that in QBC, the channel matrix $\boldsymbol{H}$ can be decomposed to two parts $\boldsymbol{H}=\boldsymbol{H}_r\boldsymbol{H}_t^{\dagger}$, where  $\boldsymbol{H}_t$ defines the probabilities for the transmitted photons to hit one of the backscatter antennas and $\boldsymbol{H}_r$ defines the probabilities for the scattered photons to be received at the receiver antennas.


Let us now assume that the backscatter antennas form an array and that there are $C$ cluttering objects near the antennas. The signal paths from the transmit antenna array to the backscatter array that go through the $C$ clutters can be expressed as:
\begin{equation}
    \boldsymbol{H}_t=\sum_{c=1}^C \sqrt{\eta_c''}e^{-i\phi_c''}\boldsymbol{e}_{N_b}(\Omega_{b,c})\boldsymbol{e}_{N_t}(\Omega_{t,c})^\dagger,
\label{eq:Ht_MIMO}
\end{equation}
where $\boldsymbol{e}_N(\Omega)^\dagger=\begin{pmatrix}1 & e^{i2\pi\Delta\Omega} &\dots  & e^{i2\pi(N-1)\Delta\Omega} \end{pmatrix}/{\sqrt{2}}$.
Similarly, the signal paths from the backscatter array to the receiver array that go through the clutters can be written as:
\begin{equation}
    \boldsymbol{H}_r=\sum_{c=1}^C \sqrt{\eta_c'''}e^{-i\phi_c'''}\boldsymbol{e}_{N_r}(\Omega_{r,c})\boldsymbol{e}_{N_b}(\Omega_{b,c})^\dagger.
 \label{eq:Hr_MIMO}
\end{equation}

If the position of the clutter objects is random, these matrices tend towards complex Gaussian random matrices with zero mean yielding Rayleigh fading. If the scattering is uncorrelated, $\boldsymbol{H}_t$ and $\boldsymbol{H}_r$ are full-rank matrices with probability one. Assuming that the number of backscatter antennas $N_b$ is less than equal to the minimum number of transmit and receive antennas, $N_b\leq \min\{N_t,N_r\}$, we have 
$r=\textrm{rank}\{\boldsymbol{H}\}=\textrm{rank}\{\boldsymbol{H}_t\}=\textrm{rank}\{\boldsymbol{H}_r\}=N_b$.

\section{Performance}\label{sec:Qreceivers}

Consider a backscatter device applying a BPSK modulation technique, and that the same symbol is applied to all the $N_b$ backscatter antennas. The modulation can be introduced in the model of the previous section by multiplying $\boldsymbol{H}$ by $x=e^{-i\varphi}$, which yelds to 
\begin{equation} \label{eq:system}
    \hat{\boldsymbol{a}}_R=x\boldsymbol{H}\hat{\boldsymbol{a}}_S + \hat{\boldsymbol{a}}_{\tilde{Z}},
\end{equation}
where $\hat{\boldsymbol{a}}_{\tilde{Z}}=\boldsymbol{US}\hat{\boldsymbol{a}}_Z$ is the receiver thermal mode vector.In the following, we focus on the operation point of QI, where $N_S=\braket{\hat{a}_S^\dagger\hat{a}_S}\ll1$, $ N_Z=\braket{\hat{a}_Z^\dagger\hat{a}_Z}\gg1$ and $\eta\ll 1$.

\subsection{SISO protocol}
In the QI setup, the entangled photon pairs of the signal (S) and the idler (I) are first generated at the TX. The S photon is transmitted from the transmit antenna and backscattered from an RF tag antenna. The receiver uses both the received S-photon  and the kept I-photon for enhancing the performance.
We consider a source able to continuously generate S-I photon pairs in the radio frequency regime in a two-mode squeezed state (TMSS) \cite{Tan08,Guha2009,Sanz17,Zhuang2017} 
\begin{equation}
\Ket{\psi}_{SI}=\sum_{n=0}^\infty \sqrt{\frac{N_S^n}{(N_S+1)^{n+1}}} \Ket{n}_S\Ket{n}_I,
\end{equation}
where $N_S$ is the average number of photons of both the signal and the idler. The parameter $N_S$ can be also interpreted as the rate at which the photons are created. The joint probability distribution of the quadratures of the TMSS is a Gaussian with zero mean value, hence the state is well defined by its covariance matrix. Indeed, if $\hat a_S$ and $\hat a_I$ represent the modes of the signal and the idler respectively, then we have that  $\langle\hat{a}_S^\dagger\hat{a}_S\rangle=\langle\hat{a}_I^\dagger\hat{a}_I\rangle=N_S$,  $\langle\hat{a}_S\hat{a}_I\rangle=\sqrt{N_S(N_S+1)}$ and $\langle\hat{a}_S^\dag\hat{a}_I\rangle=0$. 
The signal is transmitted, while the idler is kept at the receiver in order to be measured jointly with the backscattered signal.

The tag modulates the transmitted unmodulated carrier, and gives as output the received signal $\hat{a}_R=\sqrt{\eta}e^{-i\phi}\hat a_S + \sqrt{1-\eta}\hat{a}_Z $, where the phase $\phi$ is known. The task of the receiver is to distinguish between $\eta=0$ and $\eta=\bar \eta>0$. This corresponds to a OOK BC scheme, but the protocol can be easily adapted to other QBCs. In~\cite{Tan08} the authors show that, in the QI operative point, the initial correlations are useful to obtain a gain in the bit error rate up to $6$~dB with respect to the case of coherent signals as input.  Two receivers achieving a $3$~dB and a $6$~dB gain has been found in Guha et al.~\cite{Guha2009} and in Zhuang et al.~\cite{Zhuang2017} respectively. The loss in the performance is traded with experimental feasibility, as the Guha-RX involves only two-mode interactions in contrast to three-mode interactions needed in the Zhuang-RX. Both protocols present a bit error rate decreasing exponentially with $M$. The first-order behaviour in $M$ of the bit error probability is given by the Chernoff bound:
\begin{equation} \label{eq:BER}
P_\beta( M)\sim \exp\left(-\beta M\right)
\end{equation}
where $\beta$ is a coefficient which depends on the transmitter-receiver system that we are adopting, and we will refer to it as signal-to-noise ratio (SNR). In the QI case, the SNR $\beta$ is $\eta N_S /2N_Z$ using the Guha-RX, and $\eta N_S /N_Z$ in the Zhuang-RX. This is in contrast to the classical protocol, that achieves $\beta=\eta N_S /4N_Z$ with a receiver based on heterodyne detection. In the BPSK case, the visibility of the target $\eta$ does not change, and the task is to discriminate between the two phase values $\phi= \bar \phi$ and $\phi=\bar \phi+\pi$ where $\bar \phi=2\pi R /\mathsf{c}$. Both Guha-RX and the Zhuang-RX can be easily adapted for this case, achieving the same gain as in the QI setup. However, achieving a small bit error rate in the QI operating point is challenging, as $\beta \ll1$ and $M$ is limited by the bandwidth. Using a multi-antenna device is thus crucial if we want to have a future application of the QBC and QI protocols in realistic experimental conditions.   





\subsection{Paired MIMO protocol}

In the following, the performance are quantified in the case of the Zhuang-RX, where $\beta=\eta N_S/N_Z$. Lanzagorta et al.~\cite{Lanzagorta2016} proposed a multi-antenna setup composed by $N_t=N_r$ parallel transmitter-receiver pairs, where each receiver performs the measurements separately. We call this protocol as paired-MIMO (P-MIMO) scheme. The simultaneous transmissions cause interference between the signal modes, which results in an effective increasing of the thermal environmental noise. In fact, the receiver antenna $m$ would face with a have Gaussian noise with an average number of photons $N_{I,m}=|\sum_{{n=1}, {n\neq m}}^{N_t} h_{mn}|^2N_S + N_Z>N_Z$. 

%
If the receiver results are combined from all $m \in \{ 1 \dots N_r\}$ receivers using the maximal-ratio combining, the bit error probability is given by~(\ref{eq:BER}) with $\beta$ replaced by
\begin{equation}\label{eq:MIMO1}
    \beta_P= \sum_{m=1}^{N_r}\frac{N_S |h_{mm}|^2}{N_{I,m}}.
\end{equation}

%

Assume that $\textrm{trace}[\boldsymbol{H}\boldsymbol{H}^\dagger]= \sum_{k=1}^r \eta_k = r N_r \eta$.
For P-MIMO we have that $|h_{mm}|^2\sim \frac{1}{N_tN_r}\sum_{k=1}^r\eta_k=\frac{r}{N_t}\eta$ and $|\sum_{n\neq m} h_{mn}|^2\sim \frac{N_t-1}{N_tN_r}\sum_{k=1}^r\eta_k =(N_t-1)\frac{r}{N_t}\eta$, hence $\beta_P$ defined in \eqref{eq:MIMO1} reads $\beta_P=\frac{N_r\frac{r}{N_t}\beta}{(N_t-1)\frac{r}{N_t}\beta+1}$ where $\beta$ is defined in \eqref{eq:BER}.  Consequently, the P-MIMO approach is equivalent to the scheme  transmitting $M_P$ modes from single antenna transmitter such that 
\begin{equation}
    \frac{M_P}{M}=\frac{N_r \frac{r}{N_t}}{(N_t-1)\frac{r}{N_t}\beta+1},
\end{equation}
In the limit $N_t=N_r\to\infty$, the fraction of virtual modes approaches $\frac{r}{r\beta+1}$. The protocol is beneficial as long as it fulfills $r>1/(1-\beta)\geq 2$, since $\beta\ll1$. If the channel is of full rank $r=N_t=N_r$, then the fraction of virtual modes approaches $\beta^{-1}$ in the $r\beta\gg1$ limit.

\subsection{MIMO eigen-channel protocol}

In this section, we introduce the pre-coder beam-splitter matrix $\mathbf{V}$ at the transmitter, and the receiver beamformer beam-splitter matrix $\boldsymbol{U}^\dagger$ at the receiver, in order to enhance the gain in the multi-antenna setup. In the two-by-two case, each of them could be realized using a single beam-splitter, but in general case the unitary matrices need to be decomposed to a mesh consisting of multiples of two-port beam-splitters as described in \cite{clements2016optimal}. The rank of the matrix $\boldsymbol{H}$ is $r=\mathrm{rank}[\boldsymbol{H}]\leq \min\{N_r,N_b,N_t\}$. Hence, we will set up $r$ parallel transmission branches and connect them to the $r$ first ports of $\boldsymbol{V}$. No photons are transmitted to the rest of the ports. That is, their input is in the vacuum state $\ket{0}$. At the receiver we use the $r$ first output ports of $\boldsymbol{U}^\dagger$, and the rest ports only contain thermal photons. We refer to this protocol as eigen-MIMO (E-MIMO) scheme. A simple E-MIMO system is illustrated in Fig. \ref{fig:system} for $N_t=N_r=2$ scenario.

The input-output relationship of the system is given by
\begin{equation}
    \hat{\boldsymbol{a}}_{R}=\boldsymbol{U}^\dagger x\boldsymbol{H}\boldsymbol{V}\hat{\boldsymbol{a}}_{S} +\boldsymbol{U}^\dagger\boldsymbol{US}{{\hat{\boldsymbol{a}}}}_Z.
\end{equation}
It follows from the SVD of $\boldsymbol{H}$ that the system can be viewed as $r$ parallel eigen-channels:
\begin{equation}
    \hat{a}_{R,m}=\sqrt{\eta_m}e^{-i\varphi}\hat{a}_S+\sqrt{1-\eta_m}\hat{a}_{Z,m},
\end{equation}
with $m=1,\dots,r$. In this way, the $r$ parallel branches can be combined without interference. 
The bit error rate can be obtained from (\ref{eq:system}) by substituting $\beta$ with
\begin{equation} \label{eq:MIMO2}
 \beta_E=\sum_{m=1}^r{\eta_{m}} N_S /N_Z=\textrm{trace}[\boldsymbol{H}\boldsymbol{H}^\dagger] N_S /N_Z.
\end{equation}
For E-MIMO, $\beta_E = r N_r \eta N_s/N_Z = r N_r\beta$ according to \eqref{eq:MIMO2}. Hence, our proposed E-MIMO system is equivalent to transmitting $M_E$ modes from a SISO transmitter such that the resultant SNR gain reads
\begin{equation}
    \frac{M_E}{M}=r N_r.
\end{equation}
The relative gain, in terms of SNR, of the E-MIMO over the P-MIMO can be expressed as
\begin{equation}
    \frac{\beta_E}{\beta_P}=(N_t-1)r\beta + N_t.
\end{equation}
Since $\beta\ll1$, the gain is approximately proportional to the number of transmit antennas $N_t$.

\subsection{Simulation Results}
In this subsection, we show the results for the deterministic and the double-Rayleigh fading scenarios. In the presence of fading, the elements of the matrices $\boldsymbol{H}_t$ and $\boldsymbol{H}_r$ are random variables such that, $\mathbb{E}\{\textrm{trace}[\boldsymbol{H}\boldsymbol{H}^\dagger]\}=r N_r \eta$. In the double-Rayleigh fading case, the number of modes needed to achieve the same bit error rate as in the deterministic case is larger. The exact bit error probability as a function of SNR is known for the classical (no QI) case \cite{salo2006impact}.



\begin{figure}[t]
	\centering
	\includegraphics[width=\columnwidth]{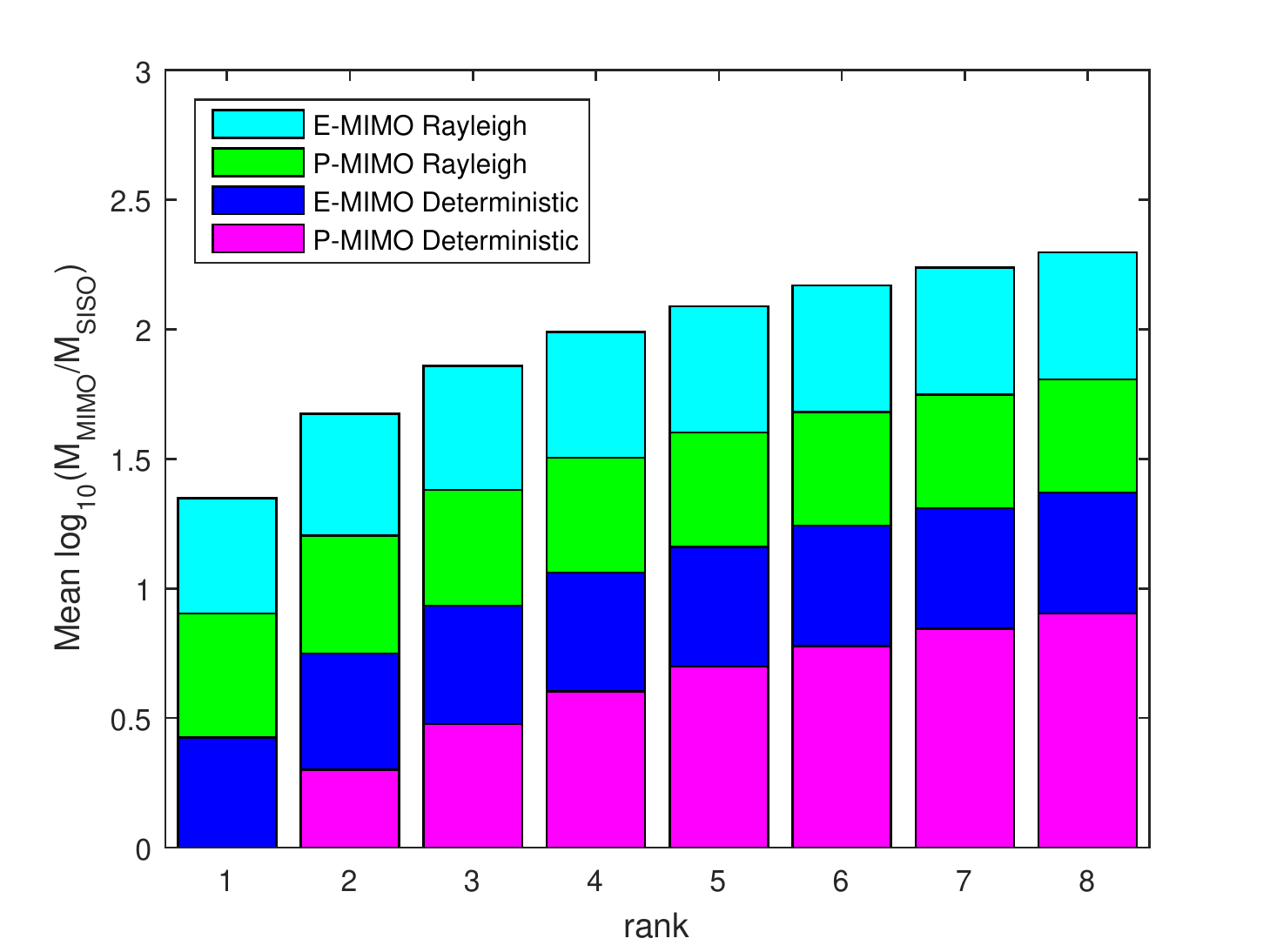} 
	\caption{$\mathbb{E}\{\log_{10}(M_P/M)\}$ and  $\mathbb{E}\{\log_{10}(M_E/M)\}$ as a function of $r$ for a MIMO system with $N_t=N_t=8$.}
	\label{fig:modes}
\end{figure}

\begin{figure}[t]
	\centering
	\includegraphics[width=\columnwidth]{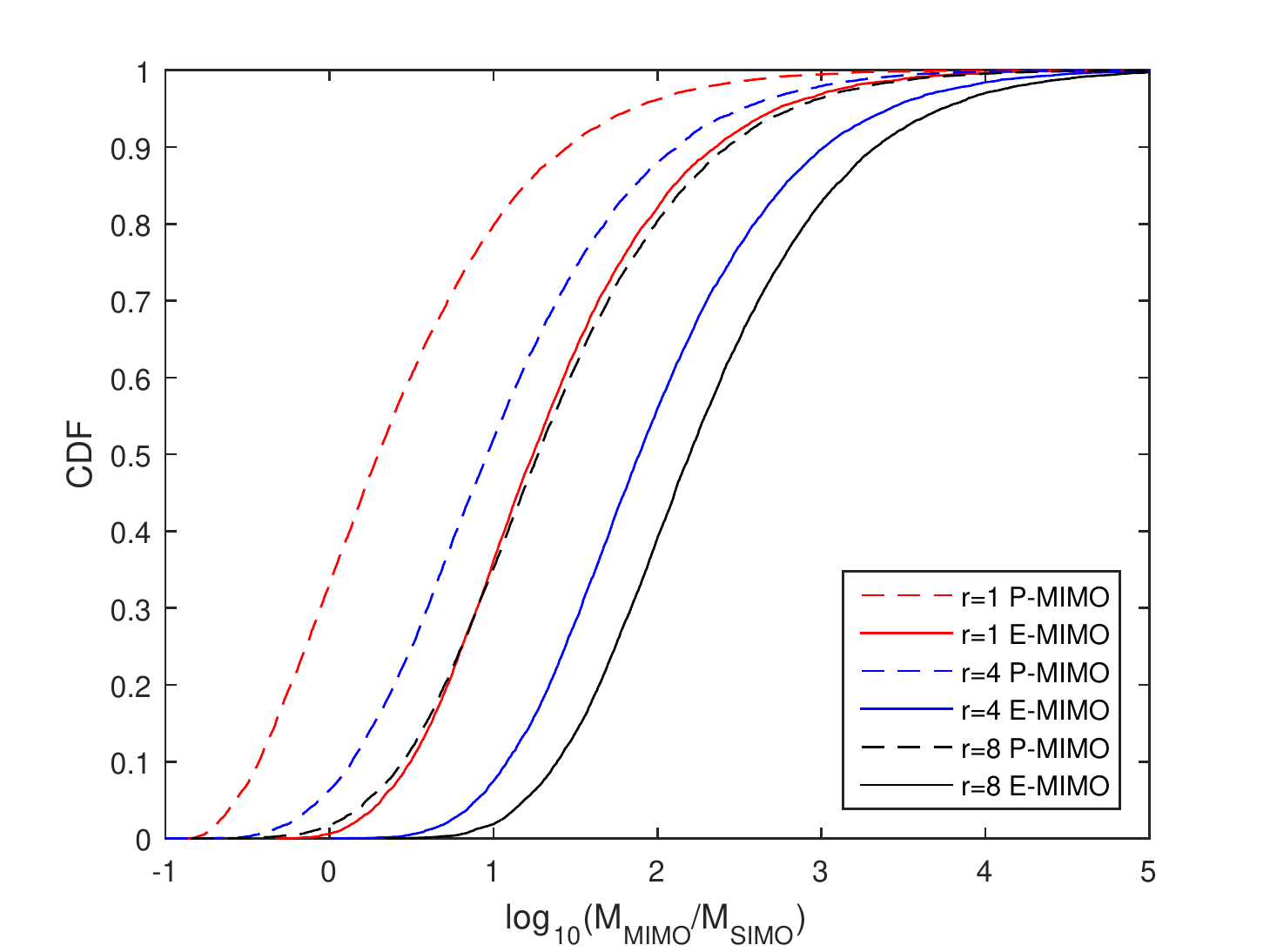} 
	\caption{CDF of $\log_{10}(M_P/M)$ and  $\log_{10}(M_E/M)$ for a MIMO system with $N_t=N_t=8$.}
	\label{fig:modesCDF}
\end{figure}

Fig.~\ref{fig:modes} shows the mean mode gain $\mathbb{E}\{\log_{10}(M_{\mathsf{MIMO})}/M_{\mathsf{SISO}}\}$ of using multi-antenna protocols with $M_{\mathsf{MIMO}}$ virtual modes over the SISO case with $M_{\mathsf{SISO}}$ modes as a function of the channel rank. Fig.~\ref{fig:modesCDF} depicts the corresponding CDFs. Fig.~\ref{fig:modes} confirms that in case of deterministic rank one channel, the P-MIMO obtains no gain over the SISO case. In the fading channel it still provides gain through averaging over the different paths. Fig.~\ref{fig:modesCDF} indicates that there exist some channel matrix realizations $\boldsymbol{H}$ where the P-MIMO fails to provide gains over single antenna E-MIMO system. It also indicates that the E-MIMO outperforms  SISO  for $r>1$ and P-MIMO for all $r$. Using multiple antennas in the tag helps to increase the channel rank $r$, and it thus increases the mode gain. The figures also show that the use of multiple antennas is an efficient way to mitigate the fading.


\section{Conclusions}
One of the limiting factor of the QI-based inference on the microwave domain is the low rate at which the entangled photon pairs can be generated. In this paper we propose a multi-antenna setup, whose channel can be decomposed as a series of beam-splitters. This allowed us to construct pre-coder beam-splitters and receiver beamforming beam-splitters such that the orthogonal eigen-channels can be accessed. Compared to the approach where multiple QI transceiver pairs are used in parallel, the eigen-channel approach significantly increase the number of distinguishable modes needed to achieve the target bit error probability. Using multiple antennas at the tag helps in increasing the number of virtual modes, and at the same time it allows to mitigate the impact of fading. The virtual modes can be utilized to compensate for the limited number of modes $M=WT$ that each quantum illumination device can generate for the given bandwidth $W$ and symbol duration $T$.


%


\ifCLASSOPTIONcaptionsoff
  \newpage
\fi


\end{document}